\documentclass{cpcauth} 
\usepackage{epsfig}

\begin{document}
\begin{frontmatter}
\title{Parallel J-W Monte Carlo Simulations of Thermal Phase Changes 
in Finite-size Systems}

\author{ R. Radev and A. Proykova}
\address{University of Sofia, Department of Atomic Physics,
5 James Bourchier Blvd. Sofia-1126, Bulgaria}

\begin{abstract}
 The thermodynamic properties of $(TeF_6)_{59}$ clusters
 that undergo temperature-driven phase
transitions have been calculated 
 with a canonical J-walking Monte Carlo technique. A parallel code for
simulations has been developed and optimized on SUN3500 and
CRAY-T3E computers.
The Lindemann criterion shows that the clusters transform from liquid to
solid and then from one solid structure to another in the temperature
region 60-130 K.
\end{abstract}


\end{frontmatter}
\section{Introduction}
The interest in phase transitions in finite systems is two-fold:
firstly, bulk properties of the material can be simulated if the system is
studied under periodic boundary conditions \cite{BINDER79} and proper 
account for rounding
and shifting of the measurable quantities is taken in the analysis.
Secondly, free finite systems, such as molecular clusters, are of interest
due to their peculiar properties, which are not observed in 
bulk systems of the same substance \cite{PROY97}. 
These are important for the synthesis of nanomaterials 
and nanodevices \cite{PROY99A}.

We use a canonical Monte Carlo (MC) method to investigate temperature-induced 
critical behavior in clusters containing 59 $TeF_6$ molecules.

Our model system contains 59 octahedral rigid $TeF_6$ molecules which 
are allowed  to rotate and translate. 
Their interaction is reliably described by a Lennard-Jones and a 
Coulomb atom-atom potential: 

\begin{eqnarray}
V({\bf q})& = & \sum\limits_{i<j}^{}{}U(r_{ij})   \\
V({\bf q})& = & \sum\limits_{i<j}^{}{}\sum\limits_{\alpha,\beta}^{}
                 4\epsilon_{\alpha\beta}\left[
                 \left( \frac{\sigma_{\alpha\beta}}{r_{ij}} \right)^{12} -
                 \left( \frac{\sigma_{\alpha\beta}}{r_{ij}} \right)^{6} \right]   + \nonumber  \\
          & + &  \sum\limits_{i<j}^{}{}\sum\limits_{\alpha,\beta}^{}
                  \left( \frac{q_{i\alpha}q_{i\beta}}{r_{ij}} \right)  \nonumber
\end{eqnarray}

\noindent where ${\bf q}$ is a generalized coordinate; $r_{ij}$ is the
distance between the $i-th$ and $j-th$ atom. The indices $\alpha,\beta$
 denote either a fluorine or a tellurium atom. The parameter values are taken
from \cite{BAR96}.

This orientation-dependent potential generates rugged potential energy surfaces.
 How to tackle the behavior of a system having a rugged
potential energy surface in MD simulations is discussed in \cite{ST01,A01}.

Here we tackle the problem with the help of the jump-walking Monte Carlo
method, \cite{FRAN90,FRAN92}, which is a smart development of the Metropolis
algorithm.
 We have parallelized the code to speed up
the computations \cite{RAD2000}.

\section{J-walking algorithm}\label{algorithm}
In its general form, the method generates trial moves from a
higher-temperature $(T_J)$
equilibrium distribution with a probability specified by
the variable $P_J$:

\begin{equation}
P = min{} \left\{ 1, q[ x^* | x ] \right\} 
\end{equation}

\noindent where the Metropolis probability function is:

\begin{equation}
q(x^*|x) = e^{ - ( \frac{1}{T} - \frac{1}{T_J} ) 
             \left\{ V(x^*) - V(x) \right\} }
\end{equation}

\noindent $V(x)$ is the potential energy 
of the system at a temperature $T$ and $V(x^*)$
is  the potential energy of a new configuration at  the temperature $T_J$.

The remaining trial moves are of conventional Metropolis 
character. 

In our study we use multi-temperature 
generalizations \cite{FREE96} of 
the basic approach.

We develop a parallel J-walking code that enables us to 
carry out a Monte Carlo simulation efficiently in a multiprocessor 
computing environment. We apply the code in the study of the thermodynamic
properties of $(TeF_6)_n$, $n=59$ clusters.

The J-walking technique can be implemented in two ways.
The first approach is to write the configurations from the simulation at the
 J-walking temperature to an external file and access these configurations 
randomly, while carrying out a simulation at the lower 
temperature. It is necessary to access the external files randomly to avoid
correlation errors \cite{FRAN95}. The large storage requirements limit the 
application of
the method to small systems. 
The second approach uses tandem walkers, one at a high temperature where 
Metropolis sampling is ergodic and multiple walkers at lower temperatures.

The best features of these two approaches can be combined into a single
J-walking algorithm \cite{METR96} with the use of multiple processors and the
Message Passing Interface (MPI) 
library. We incorporate MPI functions into the MC code to send 
and receive configuration geometries and potential energies of the clusters.
Instead  of generating external distributions and storing them before 
the actual simulation, we generate the required distributions during 
simulation and pass them to the lower-temperature walkers.

\noindent Parallel J-walking algorithm:

\begin{itemize}
\item Step 1. For each {\bf t} make Metropolis MC steps:
        \begin{itemize}
   	\item Rotate each molecule.
   	\item Translate each molecule.
   	\item Reject or accept step.
        \item Go to step 1.
        \end{itemize}

\item Step 2. After $S_1$ steps - collect statistics:
        \begin{itemize}
   	\item Potential energy histogram.
 	\item Energy average and deviation.
	\item Heat capacity $C_V$.
        \item Save current configuration.
        \end{itemize}

\item Step 3. After $S_2$ steps - make jump-walking step by exchanging the
         configurations using MPI.

\item Step 4. Go to step 1.
\end{itemize}

\begin{figure}[ht]
\epsfig{file=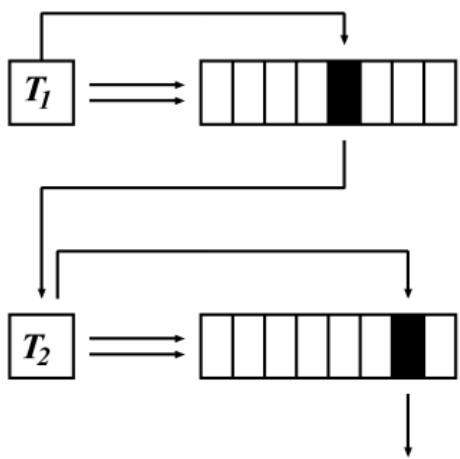}
\caption{Diagram of a simple J-walking process. $T_1$ and $T_2$ present 
the systems at two different temperatures}
\label{jwalk1}
\end{figure}

\begin{figure}[ht]
\epsfig{file=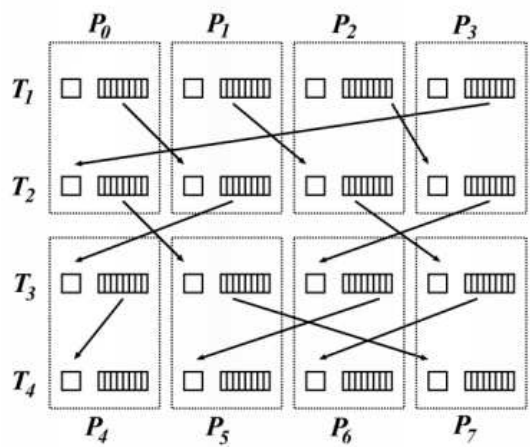}
\caption{Diagram of a parallel decomposition of data and calculation 
between processors in J-walking algorithm. In this example we have
 3 different
systems at temperatures $T_1$, $T_2$, $T_3$ and $T_4$ distributed between
6 processors}
\label{jwalk2}
\end{figure}

\noindent The diagram of model J-walking is shown in fig.\ref{jwalk1}. 
Each square 
is a Metropolis
MC simulation at a particular temperature. The set of boxes on the right-hand
 side
represents the array of previous configurations of the system, which 
are stored in the memory
to avoid correlations between the lower and the higher temperature.
At each trial jump we randomly choose one of the 4 systems.
When a configuration is transmitted to a lower-temperature process, it is 
a configuration randomly chosen from the array of higher-temperature walkers.
The current configuration of the walker then replaces the configuration just
passed from the array to another temperature.
In fig.\ref{jwalk2} we show the parallel decomposition of computation.
Each process computes part of one of 4 multistage J-walking chains, and exchanges 
configurations and energy with others.

We use array sizes of 2500 configurations.
The number of configurations is limited by the processor RAM used 
in simulation. In the parallelization implemented in our code, 
the arrays are small and 
do not inhibit applications of the method to large systems.
For the computations in the present work the number of MC passes for 
each temperature is $6.10^5$ for each cluster containing 59 
$TeF_6$ molecules (413 atoms).
We make J-walk jump attempts at every 50 Metropolis MC steps during the 
thermalisation and at every 150 steps during the main computation.

The computer code has been 
ported, tested and optimized on SUN 3500 and CRAY-T3E machines.
In our program dynamic memory management has been implemented
for optimal usage of
the memory. We find that memory and performance requirement make CRAY-T3E
more suitable for such computations.
In our runs we use 64 processors each with 64 MB RAM.
Each run of $6.10^5$ steps takes approximately 11h per CPU.

\section{Results and Conclusions}\label{results}
Using the parallel code described in the previous section we make 
sets of production runs for 59 molecule clusters.
The previous MD analysis of the temperature behavior of $TeF_6$ 
clusters pointed out a two-step structural transformation process
\cite{RAD98} from an orientationally disordered bcc structure 
below cluster solidification to an orientationally oriented ordered
monoclinic structure detected at low $T$ (below $20 K$).
The first step involves lattice 
reconstruction (bcc to monoclinic) and a partial order of one of the 
molecular axes, when the cluster is cooled down to its freezing point.
This transition has been proved to be a first-order phase change 
\cite{DAY} by detecting coexisting phases. A further temperature
decrease causes complete orientational order of the three molecular 
axes. This transformation is continuous. The diagnostic method 
developed in 
\cite{RAD98} has been implemented to animate the solid-solid 
transformations \cite{IJC}.

\begin{figure}[ht]
\epsfig{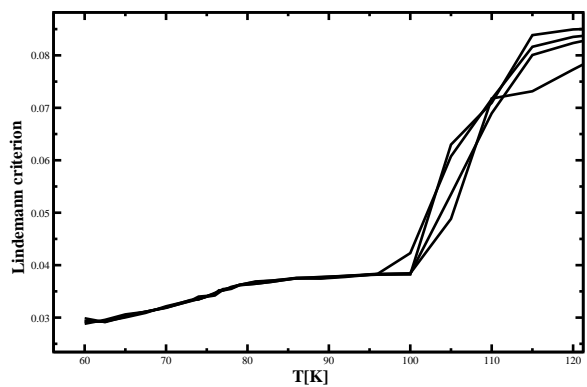}
\caption{Lindemann criterion calculated for 4 clusters of 59 molecules.}
\label{lindeman}
\end{figure}

To distinguish between the different phases, which the clusters adopt
at different temperatures, we have computed the Lindemann index \cite{LIND}:

\begin{equation}
\delta_{lin} = \frac{2}{N(N-1)} 
\sum\limits_{i,j(>i)=1}^{N}{\frac{\sqrt{<r_{ij}^{2}> - <r_{ij}>^{2}}}{<r_{ij}>}}
\label{eq_lind}
\end{equation}

\noindent where $r_{ij}$ is the distance between the centers 
of the molecules ($Te$ atoms).
For the values of $\delta_{lin} > 0.08$ the system is in liquid phase
and for $\delta_{lin} < 0.08$ is in the solid phase.
Fig.\ref{lindeman} presents the Lindemann criterion for the 4 systems.
The solidification occurs in interval $100 K - 115 K$.
In the interval $75 K - 80 K$ the angle of the curve has changed.

\begin{figure}[ht]
\epsfig{file=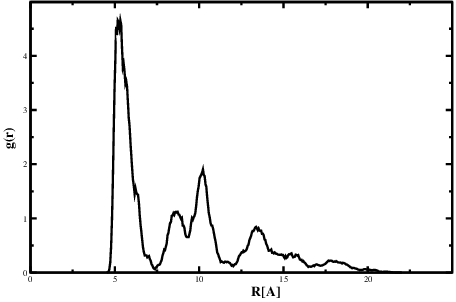}
\caption{Radial distribution function for 59 molecule $TeF_6$ cluster at 
$76 K$.}
\label{dist76}
\end{figure}

We calculate the  radial distribution function to 
reconstruct the crystal lattice.
Fig.\ref{dist157} and fig.\ref{dist76} show the 
normalized radial distribution function for distances between 
centers of the $TeF_6$ molecules ($Te$ atoms).
Fig.\ref{dist157} is at $157$. The peaks of the curve correspond 
to disordered liquid structure. Fig.\ref{dist76} is at $76 K$ where
the peaks correspond to an ordered structure (bcc lattice).

In this study a J-walking Monte Carlo algorithm has been 
implemented to study $(TeF_6)$ clusters.
A parallel J-walking code has been developed. 
Thermodynamic properties 
have been observed in the range from $60 K$ to $150 K$.
We can compare the results to the results obtained from MD simulations.

{\bf Acknowledgments}

The work was supported under the EC (contract No. HPRI-CT-1999-00026)
TRACS-EPCC.


\begin{figure}[ht]
\epsfig{file=gr-157.eps}
\caption{Radial distribution function for 59 molecule $TeF_6$ cluster at
$157 K$.}
\label{dist157}
\end{figure}

\end{document}